\documentclass[a4paper,pre,superscriptaddress,floatfix,nofootinbib,twocolumn]{revtex4-1}

\usepackage{bbm}
\usepackage[pdftex]{graphicx}
\usepackage{latexsym,amsmath,verbatim,amssymb,txfonts, mathtools}
\usepackage{color}
\usepackage{rotating}
\usepackage{verbatim}
\usepackage{multirow}
\usepackage[english]{babel}
\usepackage{comment}
\usepackage{hyperref}
\usepackage[normalem]{ulem}
\usepackage{tabularx}

\newcommand{\change}[1]{{#1}}

\begin{document}

\title{Detectability threshold in weighted modular networks}

\author{Filippo Radicchi}
\affiliation{Center for Complex Networks and Systems Research, Luddy School
  of Informatics, Computing, and Engineering, Indiana University, Bloomington,
  Indiana 47408, USA}
  \email{f.radicchi@gmail.com}

\author{Filipi N. Silva}
\affiliation{Center for Complex Networks and Systems Research, Luddy School
  of Informatics, Computing, and Engineering, Indiana University, Bloomington,
  Indiana 47408, USA}

\author{Alessandro Flammini}
\affiliation{Center for Complex Networks and Systems Research, Luddy School
  of Informatics, Computing, and Engineering, Indiana University, Bloomington,
  Indiana 47408, USA}

\author{Santo Fortunato}
\affiliation{Center for Complex Networks and Systems Research, Luddy School
  of Informatics, Computing, and Engineering, Indiana University, Bloomington,
  Indiana 47408, USA}

\author{Sadamori Kojaku}
\affiliation{School of Systems Science and Industrial Engineering, Binghamton University, Binghamton, New York 13902, USA}
\email{skojaku@binghamton.edu}

\begin{abstract}
We study the necessary condition to detect, by means of spectral modularity optimization, the ground-truth partition in networks
generated according to the weighted planted-partition model with two equally sized communities.
We analytically derive a general expression for the maximum level of mixing tolerated by the algorithm to retrieve community structure, showing that
the value of this detectability threshold depends on the first two moments of the distributions of node degree and edge weight.
We focus on the standard case of Poisson-distributed node degrees and compare the detectability thresholds of
five edge-weight distributions: Dirac, Poisson, exponential, geometric, and signed Bernoulli.
We show that Dirac distributed weights yield the smallest detectability threshold, while exponentially distributed weights increase the threshold by a factor of $\sqrt{2}$, with other distributions exhibiting distinct behaviors that depend, either or both, on the average values of the degree and weight distributions.
Our results indicate that larger variability in edge weights can make communities less detectable.
In cases where edge weights carry no information about community structure, incorporating edge weights in community detection is detrimental.
\end{abstract}

\maketitle

\section*{Introduction}

Many real networks display a mesoscopic organization
in modules or clusters often referred to as communities~\cite{fortunato2010community}. A
community is roughly defined as a subgroup of nodes with a
density of within-community connections
larger than the density of cross-community links.
The relative number of cross-cluster edges determines the level of mixing of the network.
If mixing is small enough, clusters are
sufficiently well separated one from the other to allow for some level of detection.
As mixing increases, communities become eventually undetectable, i.e., their detection
can not be performed better than random guessing.
The maximum level of mixing for the clusters to be detectable is the so-called
detectability threshold~\cite{reichardt2008detectable, decelle2011inference}.
The existence of two regimes of detectability of network modular structure
has been proven for several types of community detection algorithms, including
spectral clustering methods~\cite{nadakuditi2012graph, radicchi2013detectability, krzakala2013spectral, chen2015universal},
modularity-maximization-based methods~\cite{radicchi2014paradox},
as well as methods based on neural networks~\cite{kojaku2024network,kawamoto2018mean,chen2019supervised}.
These findings are backed up by
formal information-theoretic derivations of the detectability threshold~\cite{abbe2015community, abbe2016crossing, abbe2018community}.
The standard setting where the problem has been investigated is the
planted-partition model (PPM)~\cite{condon2001algorithms}
with two communities of equal size,
where within- and cross-cluster degrees of individual nodes are random variates extracted
from Poisson distributions with average
values respectively equal to $\langle k_{\text{in}} \rangle$ and $\langle k_{\text{out}} \rangle$.
One would naively expect communities to be detectable as long as
$\langle k_{\text{in}} \rangle - \langle k_{\text{out}} \rangle > 0$; instead, better-than-random
detection is possible only
if $\langle k_{\text{in}} \rangle - \langle k_{\text{out}} \rangle > \sqrt{\langle k_{\text{in}} \rangle + \langle k_{\text{out}} \rangle}$~\cite{decelle2011inference}.
The problem of community detectability has been studied also under other settings, for example PPMs with different types of degree distributions~\cite{radicchi2013detectability},
heterogeneous community sizes~\cite{decelle2011inference, abbe2016crossing}, non-null levels of correlation among the within- and cross-cluster degrees~\cite{radicchi2014paradox}
and hierarchical community structure~\cite{peel2024detectability}, just to mention a few of them.
The general message from all these investigations is that the value of the detectability threshold
may dramatically change from network to network, however, there is no algorithm that can
perform better than random guessing when the level of mixing is greater than the
network-specific threshold value.

The goal of the present paper is to
generalize the problem of community detectability to networks with weighted edges.
The fact that this specific problem has not been considered so far is somehow surprising given that community detection
in weighted graphs has been the focus on a large body of papers, e.g., Refs.~\cite{newman2004analysis,alves2007unveiling,lancichinetti2009benchmarks,heimo2008detecting,fan2007accuracy,peixoto2018nonparametric,macmahon2015community,tian2024spreading}.
Intuitively, one would expect 
that incorporating edge weights into community detection should enhance performance by providing richer information beyond mere topological structure~\cite{kovacs2024iterative,lu2018adaptive,berry2011tolerating,de_meo2013enhancing}. 
\change{Our analysis reveals that performance enhancement arises only when edge weights are positively correlated with community structure. If edge weights are generated independently from community structure instead, weights act as noise that impedes the proper detection of the communities. In this case, ignoring edge weights widens the detectable regime.}
These findings suggest that the utility of edge weights in community detection depends critically on whether they reinforce the community structure or not.

We derive a general detectability limit that depends on the first two moments of the node degree and the edge weight distributions, providing a unified framework for understanding community detection in weighted networks. We then focus on two specific cases: communities manifested solely through the topology with homogeneous weights, and communities manifested through edge weights with homogeneous topology. These two cases represent complementary scenarios for how community structure can emerge in weighted networks.

For both specific cases, we consider a variant of the PPM model where within- and cross-cluster degrees obey Poisson distributions with average values $\langle k_{\text{in}} \rangle$ and $\langle k_{\text{out}} \rangle$, respectively, and edges are weighted by random variates with average values $\langle w_\text{in} \rangle$ and $\langle w_\text{out} \rangle$, respectively.
We derive closed-form expressions for the detectability thresholds across five weight distributions: Dirac, Poisson, geometric, signed Bernoulli, and exponential. Our key findings reveal a hierarchy of detectability thresholds.
The Dirac distribution yields the smallest detectability threshold, matching the unweighted PPM limit and serving as the optimal baseline. Exponential weights consistently increase the threshold by a factor of $\sqrt{2}$ compared to Dirac weights. Poisson and signed-Bernoulli distributions exhibit variable behavior: they produce the largest thresholds for small total weights but approach the Dirac limit as the total weights increase. Geometric weights show intermediate behavior, with thresholds that increase and approach the exponential distribution performance as the total weights increase.

These results apply to both community manifestation scenarios—through topology with homogeneous weights and through edge weights with homogeneous topology—and are derived using spectral optimization of the modularity function. We validate our theoretical predictions by extending the analysis to multi-community networks using the Leiden algorithm, which confirms the same hierarchy of weight distributions across different numbers of communities
\change{(Appendix~\ref{sec:more_communities})}. 
We believe these findings extend to arbitrary community detection methods in weighted networks.

\section*{Methods}

\subsection*{Weighted planted-partition model}

We consider networks generated according to a planted-partition model (PPM) with two blocks containing $N$ nodes each~\cite{condon2001algorithms}. We refer to these blocks as ground-truth communities or clusters. Without loss of generality, we label with $i=1, \ldots, N$ the nodes in the first community, and with $i=N+1, \ldots, 2N$ the nodes in the second community.
Weighted connections among pairs of nodes are generated with a two-step procedure~\cite{aicher2014learning,peixoto2018nonparametric}:

\begin{enumerate}


    \item We consider each pair of distinct nodes $i$ and $j$ and associate to it a Bernoulli random variate $a_{ij}$
    with success probability $p_\text{in}$ if $i$ and $j$ are in the same community, or $p_\text{out}$ if $i$ and $j$ are in different communities.

    \item
     We consider each pair of distinct nodes $i$ and $j$ and
    associate to it a weight $w_{ij}$ extracted from the probability distribution ${\cal W}_\text{in}$ if $i$ and $j$ are in the same community, or extracted from the probability distribution ${\cal W}_\text{out}$ if $i$ and $j$ are in different communities.
\end{enumerate}

Given a network generated according to the above procedure, the degree of node $i$ is defined as
\begin{equation}
    k^{(i)} = \sum_{j=1}^{2N} a_{ij} \; .
    \label{eq:degree}
\end{equation}
If $i$ belongs to the first community, then its within-community degree is
\begin{equation}
    k^{(i)}_\text{in} = \sum_{j=1}^{N} a_{ij}
    \label{eq:in_degree}
\end{equation}
whereas its between-community degree is
\begin{equation}
    k^{(i)}_\text{out} = \sum_{j=N+1}^{2N} a_{ij} \; .
    \label{eq:out_degree}
\end{equation}
Similar expressions are valid for $i$ belonging to the second community.
In particular, we denote with ${\cal K}_\text{in}$ and ${\cal K}_\text{out}$ the resulting probability distributions
of the within-cluster or internal degree $k_\text{in}$  and the cross-cluster or external degree $k_\text{out}$, respectively.

The strength of node $i$ is defined as
\begin{equation}
    s^{(i)} = \sum_{j=1}^{2N} a_{ij} w_{ij} \; .
    \label{eq:strength}
\end{equation}
Also in this case, if node $i$ belongs to the first community, its within- and between-cluster strengths are
\begin{equation}
    s^{(i)}_\text{in} = \sum_{j=1}^{N} a_{ij} w_{ij}
    \label{eq:in_strength}
\end{equation}
and
\begin{equation}
    s^{(i)}_\text{out} = \sum_{j=N+1}^{2N} a_{ij} w_{ij} \; ,
    \label{eq:out_strength}
\end{equation}
respectively. Analogous expressions can be written if node $i$ is in the second community.

As defined in Ref.~\cite{newman2006modularity},
the generic entry of the modularity matrix of the graph is defined as
\begin{equation}
q_{ij} = a_{ij} w_{ij} - \frac{s^{(i)} s^{(j)}}{ \sum_{r=1}^{2N} s^{(r)}} \; .
    \label{eq:modularity}
\end{equation}
On the right hand side of the equation,
the second term represents the expected value of the first term in a random graph.


\subsection*{Detectability threshold}

We focus our attention on the best bi-partition that can be recovered by means of spectral modularity optimization~\cite{newman2006modularity}. Spectral methods are among the most widely used approaches for community detection; they also have the advantage of allowing for analytical derivation of detectability thresholds. Moreover, spectral modularity maximization achieves optimal performance in terms of the detectability limit for unweighted networks as long as the average degree is large enough, making it an ideal theoretical framework for extending to weighted networks. Community structure in our PPM composed of only two communities can be retrieved by looking at the principal eigen-pair $(\lambda, \mathbf{v})$ of the modularity matrix of Eq.~\eqref{eq:modularity}. We quantify the ability of the method to retrieve the ground-truth community structure in instances of the weighted PPM using the order parameter
\begin{equation}
P = \frac{1}{\sqrt{2N}}  \left|\sum_{i=1}^{N} v_i \right|  + \frac{1}{\sqrt{2N}}  \left|\sum_{i=N+1}^{2N} v_i \right|   \; .
    \label{eq:order}
\end{equation}
The order parameter $P$ is the normalized sum of the within-cluster components of the principal eigenvector; we expect these components to have coherent signs when clusters are detectable, and random signs otherwise~\cite{newman2006modularity}. The pre-factor $1/\sqrt{2N}$ arises from the normalization of the eigenvector $\mathbf{v}$ and serves to set $0 \leq P \leq 1$. \change{$P > 0$ denotes that the method can, at least partially, retrieve the ground-truth partition of the network; $P \simeq 0$ indicates that clusters are not detectable.}

In the detectability regime, the largest eigenvalue $\lambda$ of the modularity matrix is shown to be well approximated by
\begin{equation}
\lambda = \frac{\langle \Delta_{s}^2 \rangle}{\langle \Delta_{s} \rangle} \; ,
\label{eq:main}
\end{equation}
where $\Delta_{s}=s_\text{in} - s_\text{out}$ is the difference between the within-cluster node strength (i.e., $s_\text{in}$)
and the cross-cluster node strength (i.e., $s_\text{out}$).
See Ref.~\cite{radicchi2013detectability} for the derivation of Eq.~(\ref{eq:main}).
The expression of Eq.~(\ref{eq:main}) is valid as long as the individual components of the principal eigenvectors are
proportional to the difference between the within- and cross-cluster strengths of the individual nodes.
In the above expression,
$\langle x \rangle$ indicates the average of the quantity $x$ over the ensemble of PPMs corresponding to a given set of model parameters. The largest eigenvalue $\lambda$ is therefore well approximated by the ratio between the second and the first moments of the distribution of $\Delta_{s}$.

As a rule of thumb, we expect that $\lambda$ assumes large values when communities are well separated one from the other; it becomes instead comparable with the largest eigenvalue of a graph with similar average strength but no block structure when communities are too mixed to be detected. The latter condition is the one typically used to determine the so-called detectability threshold, i.e., the maximum level of mixing that the spectral method can tolerate in order make a prediction better than random guessing~\cite{nadakuditi2012graph}. When the network is unweighted, although the above procedure assumes that a spectral method  is used for the detection of communities, the actual threshold that can be derived from this analysis matches exactly the one obtained with a purely information-theoretical approach~\cite{abbe2015community}. In this respect, it represents a limit that can not be overcome by any community detection method. We believe that the equivalence between the spectral-modularity-based  and the information-theoretic thresholds extends also to weighted networks.

\subsection*{Moments of the strength and the strength-differential distributions}

To derive detectability thresholds, we need to express the first two moments $\langle \Delta_{s} \rangle$ and $\langle \Delta_{s}^2 \rangle$ of the strength-difference distribution ${\cal D}$
as functions of the moments of the degree and weight distributions $\cal K_\text{in}, \cal K_\text{out}, \cal W_\text{in} $ and $\cal W_\text{out} $.
Let ${\cal S}_\text{in}(s_\text{in})$ be the probability that a node has within-cluster strength $s_\text{in}$, and ${\cal S}_\text{out}(s_\text{out})$ similarly for cross-cluster strength.
The probability that $\Delta_{s} = s$ is given by the convolution
\begin{equation}
    {\cal D}(s) = \sum_{s_\text{in} - s_\text{out} = s} \,  {\cal S}_\text{in} (s_\text{in}) \, {\cal S}_\text{out} (s_\text{out}) \;,
\label{eq:conv2}
\end{equation}
if the weights are discrete random variables. For real-valued weights, the summation must be replaced by the integral ${\cal D}(s) = \int\int \,  {\cal S}_\text{in} (s_\text{in}) \, {\cal S}_\text{out} (s_\text{out}) \delta(s_\text{in} - s_\text{out}-s){\rm d}s_{\text{in}}{\rm d}s_{\text{out}}$.
The derivation of our results is insensitive to whether weights are discrete or continuous.
A node's within-cluster strength is the sum of weights on its $k$ within-cluster edges, where $k$ is drawn from ${\cal K}_\text{in}$ and each edge weight is independently drawn from ${\cal W}_\text{in}$.
This leads to
\begin{equation}
    {\cal S}_\text{in} (s) =
    \sum_{k} {\cal K}_\text{in}(k) \sum_{w_1 + \ldots + w_k = s} {\cal W}_\text{in}(w_1) {\cal W}_\text{in}(w_2) \cdots {\cal W}_\text{in}(w_k) \; .
    \label{eq:conv1}
\end{equation}
A similar expression holds for the distribution ${\cal S}_\text{out}$.

The strength difference ${\cal D}(s)$ is a convolution of the distributions ${\cal S}_\text{in}$ and ${\cal S}_\text{out}$.
As such, it is convenient to introduce the moment generating function (MGF) of ${\cal D}(s)$ to derive the first two moments of ${\cal D}(s)$.
The MGF of a distribution $f(k)$ is defined as
\[
M_f(x) = \langle e^{x k} \rangle \; ,
\]
where $\langle \cdot \rangle$ denotes the expectation value over the distribution $f(k)$.
The MGF $M_{\cal D}(x)$ of the distribution ${\cal D}$ is
\[
M_{\cal D}(x) = \langle e^{x (s_\text{in} - s_\text{out})} \rangle =
M_{{\cal S}_\text{in}}(x) \, M_{{\cal S}_\text{out}}(-x) \; ,
\]
where we use the fact that $s_\text{in}$ and $s_\text{out}$ are independent random variables.
The MGF $M_{{\cal S}_\text{in}}(x)$ of the distribution ${\cal S}_\text{in}$ is
\[
M_{{\cal S}_\text{in}}(x) =
\sum_{k} {\cal K}_\text{in}(k) \left[ M_{{\cal W}_\text{in}}(x) \right]^k \; ,
\]
which follows from the fact that for a node with degree $k$ drawn from ${\cal K}_\text{in}$, the strength is the sum of $k$ independent and identically distributed edge weights. Similarly, we have
\[
M_{{\cal S}_\text{out}}(x) =
\sum_{k} {\cal K}_\text{out}(k) \left[ M_{{\cal W}_\text{out}}(x) \right]^k \; .
\]
By taking the first and second derivative of $M_{{\cal D}}(x)$ and evaluating them at $x=0$, we obtain the first and second moments of the distribution ${\cal D}$ (see Appendix~\ref{sec:moment_calculations} for detailed calculations):
\begin{equation}
\langle \Delta_{s} \rangle   = \langle s_\text{in} \rangle   - \langle s_\text{out} \rangle
\label{eq:mD1}
\end{equation}
and
\begin{equation}
\langle \Delta_{s}^2 \rangle  = \langle s^2_\text{in} \rangle \,
- 2 \langle s_\text{in} \rangle \langle s_\text{out} \rangle + \langle s^2_\text{out} \rangle \; .
\label{eq:mD2}
\end{equation}
where the individual strength moments are:
\begin{equation}
    \langle s_\text{in} \rangle = \langle w_\text{in} \rangle \langle k_{\text{in}} \rangle
    \label{eq:mS1}
\end{equation}
and
\begin{equation}
\langle s^2_\text{in} \rangle = \langle w^2_\text{in} \rangle \langle k_{\text{in}} \rangle + \langle w_\text{in} \rangle^2 \left( \langle k^2_\text{in} \rangle - \langle k_{\text{in}} \rangle  \right),
\label{eq:mS2}
\end{equation}
with similar expressions for $\langle s_\text{out} \rangle$ and $\langle s^2_\text{out} \rangle$.
Using Eqs.~(\ref{eq:mD1}),~(\ref{eq:mD2}),~(\ref{eq:mS1}) and~(\ref{eq:mS2}), we can estimate the largest eigenvalue of the modularity matrix, i.e., Eq.~(\ref{eq:main}), as a function of the first two moments of the input distributions ${\cal K}_\text{in}$, ${\cal K}_\text{out}$, ${\cal W}_\text{in}$ and ${\cal W}_\text{out}$.

\section*{Results}

Having established the theoretical framework for detectability in weighted networks, we now apply our methods to derive explicit detectability thresholds for five fundamental weight distributions: Dirac, Poisson, exponential, signed Bernoulli and geometric.
These distributions model different ways of assigning weights to edges in real-world networks.
As mentioned earlier,
we consider the PPM topology where the degree distributions follow the binomial distributions
\begin{equation}
{\cal K}_\text{in}(k) = {N-1 \choose k} \, p_{\text{in}}^{k} \left(1 - p_\text{in} \right)^{N-1-k}   \;
    \label{eq:standard1}
\end{equation}
and
\begin{equation}
{\cal K}_\text{out}(k) = {N \choose k} \, p_{\text{out}}^{k} \left(1 - p_\text{out} \right)^{N-k}  \; .
    \label{eq:standard2}
\end{equation}
Assuming $N \gg 1$,
we have
\begin{equation}
  \langle k_{\text{in}} \rangle = N p_{\text{in}} \; ,
\label{eq:kappa_in}
\end{equation}
and
\begin{equation}
\langle k_{\text{in}}^2 \rangle =
\langle k_{\text{in}} \rangle^2 + \langle k_{\text{in}} \rangle \; ,
\label{eq:kappa_in_2}
\end{equation}
thus we can reduce Eq.~(\ref{eq:mS2}) to
\begin{equation}
\langle s_\text{in}^2 \rangle = \langle w^2_\text{in} \rangle
\langle k_{\text{in}} \rangle
+ \langle w_\text{in} \rangle^2
\langle k_{\text{in}} \rangle^2
\; .
 \label{eq:mS2a}
\end{equation}

Similar expressions are valid for $\langle s_\text{out}^2\rangle$, which are defined
on the basis of the first moment of the cross-community degree distribution
\begin{equation}
  \langle k_{\text{out}} \rangle = N p_{\text{out}} \; .
\label{eq:kappa_out}
\end{equation}

As detailed in Appendix~\ref{sec:derivation_of_lambda}, the principal eigenvalue of the modularity matrix of the weighted PPM reads
\begin{equation}
\lambda = \langle k_{\text{in}} \rangle \langle w_\text{in} \rangle - \langle k_{\text{out}} \rangle \langle w_\text{out} \rangle + \frac{\langle k_{\text{in}} \rangle \langle w^2_\text{in} \rangle + \langle k_{\text{out}} \rangle \langle w^2_\text{out} \rangle}{\langle k_{\text{in}} \rangle \langle w_\text{in} \rangle - \langle k_{\text{out}} \rangle \langle w_\text{out} \rangle} \; .
\label{eq:main_equation}
\end{equation}

To derive a unified framework for different weight distributions, we parameterize the second moment of the edge weight
distributions as a second-order polynomial of its mean value:
\begin{equation}
\langle w^2_\text{in} \rangle =
\alpha_0 + \alpha_1 \langle w_\text{in} \rangle + \alpha_2 \langle w_\text{in} \rangle^2
\; .
\label{eq:second_order_polynomial}
\end{equation}
The same parameterization is valid for the cross-community weights:
\begin{equation}
\langle w^2_\text{out} \rangle
= \alpha_0 + \alpha_1 \langle w_\text{out} \rangle + \alpha_2 \langle w_\text{out} \rangle^2 \; .
\label{eq:second_order_polynomial_out}
\end{equation}
This parameterization captures the variance structure of several important distributions through different values of $(\alpha_0, \alpha_1, \alpha_2)$, i.e.,
\begin{equation}
    (\alpha_0, \alpha_1, \alpha_2) =
    \left\{
    \begin{array}{ll}
        (0, 0, 1) & \text{(Dirac)} \\
        (0, 1, 1) & \text{(Poisson)} \\
        (0, -1, 2) & \text{(Geometric)} \\
        (0, 0, 2) & \text{(Exponential)} \\
        (1, 0, 0) & \text{(Signed Bernoulli)}
    \end{array}
    \right.
    \label{eq:second_order_polynomial_family}
\end{equation}

We begin by examining the case where edge weights are homogeneous across communities, i.e.,
$\langle w_\text{in} \rangle = \langle w_\text{out} \rangle$.
We specify the scale of the edge weights by
\begin{equation}
W = \langle w_\text{in} \rangle + \langle w_\text{out} \rangle.
\label{eq:omega}
\end{equation}
In this scenario, community structure manifests only through the network topology, not through weight heterogeneity. This provides a crucial benchmark to validate our results against the well-established results in the literature.

Substituting our expressions for the moments of the weight distributions (Eqs.~\eqref{eq:mS1}--\eqref{eq:second_order_polynomial_out}) into the eigenvalue formula
(Eq.~\eqref{eq:main_equation}),
we obtain the following key result (see Appendix~\ref{sec:derivation_of_lambda} for detailed derivation):
\begin{equation}
    \lambda = \frac{W}{2}\Delta_{k} + \frac{KC}{2W}\cdot\frac{1}{\Delta_{k}}
\label{eq:lambda_moment_homogeneous_weights}
\end{equation}
where we have defined
\begin{equation}
K = \langle k_{\text{in}} \rangle + \langle k_{\text{out}} \rangle \; ,
\label{eq:kappa}
\end{equation}
\begin{equation}
\Delta_{k} = \langle k_{\text{in}} \rangle - \langle k_{\text{out}} \rangle
\label{eq:delta}
\end{equation}
and
\begin{equation}
    C =  4\alpha_0  + 2W\alpha_1 + W^2\alpha_2 \; .
    \label{eq:c_value}
\end{equation}

For fixed $K$, $\lambda$ is the sum of the first term linear in $\Delta_k$ and the second term inversely proportional to $\Delta_k$ (Eq.~\eqref{eq:lambda_moment_homogeneous_weights}). As $\Delta_k$ decreases from the maximum value of $\Delta_k=K$, $\lambda$ decreases because the first term dominates in $\lambda$, meaning that the value of the modularity function associated with the ground-truth partition of the network decreases as the community structure becomes mixed.
However, as $\Delta_k$ keeps decreasing, the second term dominates in $\lambda$, and $\lambda$ increases again. Such an increase does not correspond to a real increase of the strength of the modular structure, rather only to the violation of the assumptions made to reach Eq.~(\ref{eq:main}). The emergence of such a physically meaningless solution occurs at the detectability threshold. To compute the latter quantity, we set the derivative of $\lambda$ with respect to $\Delta_k$ equal to zero, i.e.,
\begin{equation}
    \left. \frac{\partial \lambda}{\partial \Delta_k} \right|_{\Delta_k = \Delta_k^*} =
    \frac{1}{2}W - \frac{1}{\Delta_{k}^2} \cdot \frac{KC}{2W}
    = 0
\end{equation}
leading to the detectability threshold
\begin{equation}
\Delta_k^* = \frac{\sqrt{K C}}{W}
\label{eq:lambda_moment_homogeneous_weights_general}
\end{equation}
Equation~(\ref{eq:lambda_moment_homogeneous_weights_general}) is a general expression that includes the classic detectability threshold for unweighted networks as a special case, as we will demonstrate in the case of the Dirac weights.

Next, we consider the complementary scenario where the network topology is homogeneous (an Erd\H{o}s--R\'enyi random graph with
$\langle k_{\text{in}} \rangle = \langle k_{\text{out}} \rangle = K/2$
), but communities are encoded solely through edge weight differences.
Defining
\begin{equation}
\Delta_{w} = \langle w_\text{in} \rangle - \langle w_\text{out} \rangle
\label{eq:delta_omega}
\end{equation}
as the difference between the average weight of with- and cross-cluster edges, we obtain:
\begin{equation}
    \lambda = \frac{\Delta_{w}}{2} \left(K + \alpha_2 \right) + \frac{C}{2 \Delta_{w}}
    \label{eq:lambda_moment_homogeneous_weights_general_omega}
\end{equation}
(see Appendix~\ref{sec:detailed_derivation_homogeneous_topology} for the derivation of this expression).
Thus, by taking the derivative of $\lambda$ with respect to $\Delta_{w}$ and setting it equal to zero, we obtain the detectability threshold
\begin{equation}
\Delta_{w}^* = \sqrt{\frac{C }{K + \alpha_{2}}}
    \label{eq:delta_omega_threshold_homogeneous_topology}
\end{equation}

\change{We notice that the eigenvalue $\lambda$ has the same physical dimensions as of the edge weights; this implies that some of the above predictions can change if the weights of the edges are measured using a different unit. More specifically, assume that we change our unit of measure so that the weight $w'_{ij} = \beta w_{ij}$ for all pairs of nodes $i$ and $j$. The use of the new unit changes the values of several of the above-defined quantities. For example, the sum of all edge weights becomes $W' = \beta W$. Similarly, we can write: $\langle w' \rangle = \beta \langle w \rangle$, $\langle (w')^2 \rangle = \beta^2 \langle w^2 \rangle$ , $\Delta_w' = \beta \Delta_w$, $\alpha_0' = \beta^2 \alpha_0$,  $\alpha_0' = \beta \alpha_1$, $\alpha_2'  = \alpha_2$ and $C' = \beta^2 C$. Plugging these expressions into Eqs.~(\ref{eq:lambda_moment_homogeneous_weights}) or ~(\ref{eq:lambda_moment_homogeneous_weights_general_omega}), we find that $\lambda' = \beta \lambda$, as one would naturally expect. More importantly, the threshold of Eq.~(\ref{eq:lambda_moment_homogeneous_weights_general}) is unaffected by the change of unit; also this results is expected as the threshold refers to the topology of the graph and not the weights of its edges. Instead, the threshold of Eq.~(\ref{eq:delta_omega_threshold_homogeneous_topology}) is rescaled by the factor $\beta$ since this quantity is measured in the same units as of the edge weights.
}

With this general framework established, we now examine specific weight distributions that represent different network scenarios. Our theoretical approach applies to weight distributions that can be characterized through their first two moments and satisfy certain regularity conditions. While we focus on commonly encountered distributions,
the framework extends to other distributions with well-defined MGFs, though the specific form of detectability thresholds may vary.

\subsection*{Dirac weights}
\label{sec:dirac_weights}

The Dirac distribution represents the simplest case where edge weights are deterministic functions of the community membership. This scenario arises in networks with fixed interaction strengths, such as power grids with predetermined transmission capacities or social networks where relationship types (e.g., friend, colleague, family) determine fixed interaction strengths. Notably, networks constructed from correlation matrices---commonly used in neuroscience, finance, and genomics---converge to this Dirac case in the limit of large sample sizes, as sample correlations approach their true values and variance vanishes (see Appendix~\ref{sec:correlation_matrix} for details).

For Dirac-distributed weights, all edges within (between) communities have identical weight
$\langle w_{\text{in}} \rangle$ ($\langle w_{\text{out}} \rangle$),
yielding zero variance. This corresponds to $\alpha_0 = 0$, $\alpha_1 = 0$, and $\alpha_2 = 1$ in our parameterization. Substituting these values into Eq.~\eqref{eq:lambda_moment_homogeneous_weights}, we get
\begin{equation}
\lambda =
\frac{W\, \Delta_k}{2}  + \frac{KW}{2 \Delta_k} \; ,
\label{eq:lambda_dirac}
\end{equation}
and Eq.~\eqref{eq:lambda_moment_homogeneous_weights_general} reads
\begin{equation}
\Delta_k^* = \frac{\sqrt{K \cdot W^2}}{W} = \sqrt{K}
\label{eq:dirac_det}
\end{equation}
This recovers the classic result for unweighted networks, validating our general framework.

For the case of homogeneous topology ($\Delta_{k} = 0$), we obtain:
\begin{equation}
\lambda = \frac{\Delta_{w}}{2} \left(K + 1 \right) + \frac{C}{2 \Delta_{w}}
\label{eq:lambda_dirac_w}
\end{equation}
\begin{equation}
\Delta_{w}^* = \frac{W}{\sqrt{K+1}}
\label{eq:dirac_w_det}
\end{equation}

\subsection*{Poisson weights}

The Poisson distribution emerges in networks where edge weights represent counts of discrete interactions or events. Examples include social networks were weights
represent co-occurrences, e.g., shared friends, shared publications, co-sponsored bills in the US Congress~\cite{liebig2016fast,saracco2017inferring}.

For the Poisson distribution,
we have
$\alpha_0 = 0$,
$\alpha_1 = 1$, and $\alpha_2 = 1$ in our parameterization. Since Poisson random variables take non-negative integer values, the parameter space requires
$\langle w_{\text{in}} \rangle, \langle w_{\text{out}} \rangle \geq 0$,
which constrains the total average weight $W \geq 0$. Applying the general formula Eq.~\eqref{eq:lambda_moment_homogeneous_weights_general}, we obtain:
\begin{equation}
\lambda = \frac{\Delta_{k}}{2}W + \frac{1}{\Delta_{k}} \cdot \frac{KC}{2W}
\label{eq:lambda_poisson}
\end{equation}
\begin{equation}
    \Delta_k^* = \sqrt{\frac{K \left( W +2\right)}{W}}
\label{eq:poisson_det}
\end{equation}
Unlike the Dirac case, the detectability threshold now depends on the total weight $W$. For small $W$, the threshold is large, making communities difficult to detect. As $W$ increases, the threshold approaches $\sqrt{K}$, converging to the Dirac result.
Notably, $\Delta_k^*$ for the Poisson distribution is larger than the Dirac case,
suggesting that the communities in the network of Poisson-distributed weights are harder to detect than in the Dirac case in terms of the detectability limit.
By contrast, the detectability threshold of Dirac distributed nodes' degrees is larger than when degrees obey a Poisson distribution~\cite{radicchi2013detectability}.

For the case of homogeneous topology ($\Delta_{k} = 0$), we obtain:
\begin{equation}
\lambda = \frac{\Delta_{w}}{2} \left(K + 1 \right) + \frac{C}{2 \Delta_{w}}
\label{eq:lambda_poisson_w}
\end{equation}
\begin{equation}
    \Delta_{w}^* = \sqrt{\frac{W (W +2)}{K + 1}}
\label{eq:poisson_w_det}
\end{equation}
which, again, is larger than the Dirac case.

\subsection*{Exponential weights}

The exponential distribution emerges in networks where edge weights represent waiting times, durations, or distances between events. Notably, the exponential distribution has maximum entropy among all continuous distributions with a fixed mean, making it the
most meaningful choice to generate random uncorrelated networks~\cite{bianconi2007entropy,bianconi2009entropy, vallarano2021fast,garlaschelli2009generalized}.

For the exponential distribution,
our parametrization reads $\alpha_0 = 0$, $\alpha_1 = 0$, and $\alpha_2 = 2$. Since exponential random variables are strictly positive, the parameter space requires
$\langle w_{\text{in}} \rangle, \langle w_{\text{out}} \rangle > 0$
which ensures $W > 0$. From Eq.~\eqref{eq:lambda_moment_homogeneous_weights_general}:
\begin{equation}
\lambda = \frac{\Delta_{k}}{2}W + \frac{1}{\Delta_{k}} \cdot \frac{KC}{2W}
\label{eq:lambda_expon}
\end{equation}
\begin{equation}
\Delta_k^* = \sqrt{2K}
\label{eq:expon_det}
\end{equation}
This threshold is exactly $\sqrt{2}$ times larger than the Dirac case, reflecting the increased difficulty in detecting communities when edge weights have high variance.
Notably, the detectability threshold does not depend on $W$ in the case of the exponential distribution, which is different from the Poisson case.

For the case of homogeneous topology ($\Delta_k = 0$), we obtain:
\begin{equation}
\lambda = \frac{\Delta_{w}}{2} \left(K + 2 \right) + \frac{C}{2 \Delta_{w}}
\label{eq:lambda_expon_w}
\end{equation}
\begin{equation}
\Delta_{w}^* = W\sqrt{\frac{2}{K+2}} \; ,
\label{eq:expon_w_det}
\end{equation}
still $\sqrt{2}$ larger than for the Poisson case as $K \to \infty$.

\subsection*{Geometric weights}

The geometric distribution arises in networks where edge weights represent the number of trials until the first success in a sequence of independent Bernoulli trials. The geometric distribution is useful for modeling over-dispersed count data, and as for the exponential distribution, it has maximum entropy among all discrete distributions with a fixed mean~\cite{bianconi2007entropy,bianconi2009entropy, garlaschelli2009generalized,vallarano2021fast}.

For the geometric distribution,
our parametrization reads
$\alpha_0 = 0$, $\alpha_1 = -1$, and $\alpha_2 = 2$.
The parameter space requires
$\langle w_{\text{in}} \rangle, \langle w_{\text{out}} \rangle \geq 1$
which constrains $W \geq 2$.
This follows from the fact that $\langle w_{\text{in}} \rangle$ and $ \langle w_{\text{out}} \rangle$ are equal to the inverse probabilities of success of independent Bernoulli trials.
We have
\begin{equation}
\lambda = \frac{W}{2} \Delta_k + \frac{K(W-1)}{\Delta_k}
\label{eq:lambda_geometric}
\end{equation}
and
\begin{equation}
\Delta_k^* = \sqrt{\frac{2K(W - 1)}{W}}.
\label{eq:geometric_det}
\end{equation}
Likewise the Poisson weights, the detectability threshold depends on the total weight $W$ and is always smaller than the exponential case, with the difference approaching zero as $W$ increases.

For homogeneous topology:
\begin{equation}
\lambda = \frac{\Delta_w (K+2)}{2} + \frac{W(W-1)}{\Delta_w}
\label{eq:lambda_geometric_w}
\end{equation}
and
\begin{equation}
\Delta_{w}^* = \sqrt{\frac{2(W - 1)W}{K+2}},
\label{eq:geometric_w_det}
\end{equation}
which also approaches to the exponential case from below in the limit of large $W$.

\subsection*{Signed-Bernoulli weights}

Signed weights represent a different scenario where edges can have both positive and negative values, commonly used for representing friend/foe relationships in social networks~\cite{altafini2012dynamics,facchetti2011computing}, positive/negative correlations~\cite{tian2024spreading,sporns2016modular}, or credit/debit relationships in financial networks~\cite{ferreira2021loss}. We consider the case where weights are signed-Bernoulli
variables, thus equal to $+1$ with probability $z_\text{in}$ or $z_\text{out}$ for within- and across-cluster edges, respectively;
otherwise their weight is $-1$. The average value of the within-cluster edge weights is
$\langle w_\text{in} \rangle = z_\text{in} - (1 - z_\text{in}) = 2 z_\text{in} - 1$. Similarly, we have
$\langle w_\text{out} \rangle = 2 z_\text{out} - 1$

For this distribution,
our parametrization yields $\alpha_0 = 1$, $\alpha_1 = 0$, and $\alpha_2 = 0$. Since sign weights take values in $\{-1, +1\}$, their mean weight
is in the range $[-1,1]$,
which implies $-2 \leq W \leq 2$. For meaningful community detection, we require $W > 0$ to ensure that the communities are assortative, i.e., the within-community and cross-community weights are likely to be positive and negative, respectively.

From Eq.~\eqref{eq:lambda_moment_homogeneous_weights_general}, we obtain:
\begin{equation}
\lambda = \frac{W}{2} \Delta_k + \frac{K}{2\Delta_k}
\label{eq:lambda_sign}
\end{equation}
and
\begin{equation}
\Delta_k^* = \frac{2\sqrt{K}}{W}
\label{eq:sign_det}
\end{equation}
This threshold increases as the sign imbalance $W$ decreases to zero, making detection harder. This echoes the result that edge weights can impede community detection  in terms of the detectability limit when they do not reflect the community structure.

For the homogeneous topology case:
\begin{equation}
\lambda = \frac{\Delta_w K}{2} + \frac{1}{2 \Delta_w}
\label{eq:lambda_sign_w}
\end{equation}
and
\begin{equation}
\Delta_{w}^* = \frac{2}{\sqrt{K}}
\label{eq:sign_w_det}
\end{equation}
Remarkably, this threshold is independent of $W$.
In other words, when the topology does not reflect the community structure but the sign of edge weights does, the sign imbalance does not underpin the detectability limit. It is the average degree of the nodes that matters.

\subsection*{Numerical validation}

Having established the theoretical predictions for the detectability thresholds of different weight distributions, we now turn to numerical experiments to validate these analytical results. Our simulations focus on finite networks with specific parameter ranges, and we acknowledge that the agreement between theory and simulation is generally good but not perfect, with some systematic deviations expected due to finite-size effects and the asymptotic nature of our theoretical framework.

\begin{figure}[!htb]
    \centering
    \includegraphics[width=0.5\textwidth]{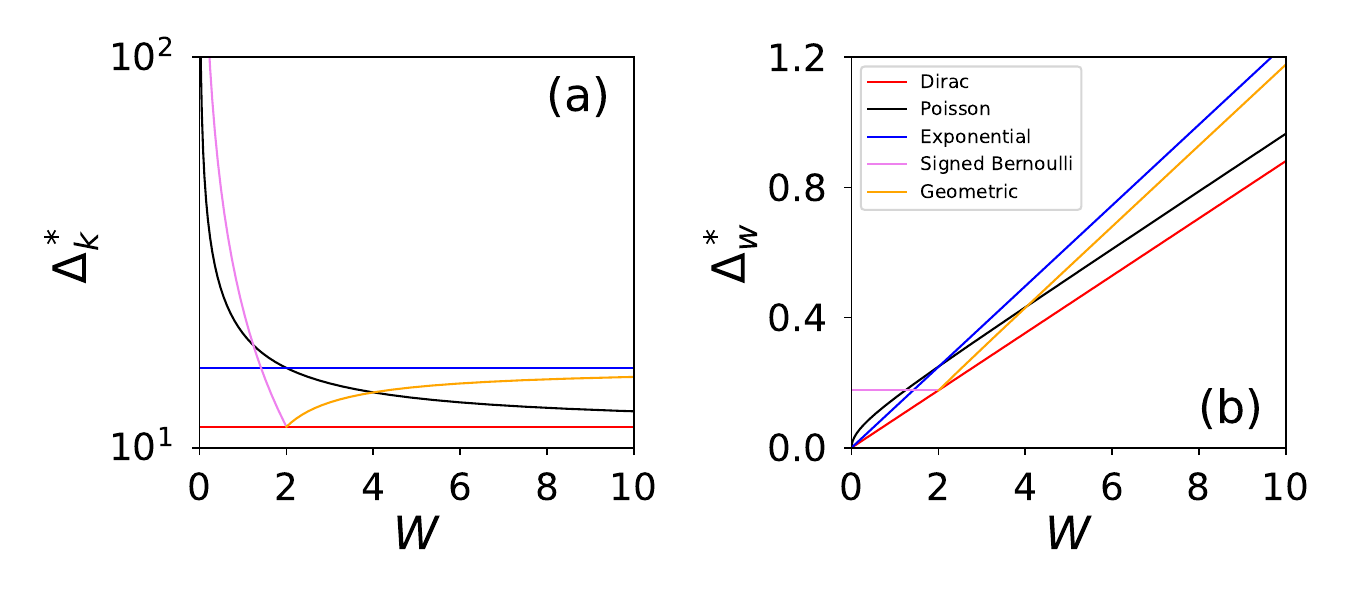}
    \caption{(a) Detectability threshold $\Delta_k^*$ for the weighted planted partition model as a function of $W$, i.e., the sum of average value of the within- and cross-community edge weights. Here $N=1024$ and $K=128$. Different curves corresponds to the different distributions of the weights we consider. Predictions are given by Eq.~(\ref{eq:dirac_det}) for the Dirac distribution, Eq.~(\ref{eq:poisson_det}) for the Poisson distribution, Eq.~(\ref{eq:expon_det}) for the exponential distribution, Eq.~(\ref{eq:geometric_det}) for the geometric distribution, and Eq.~(\ref{eq:sign_det}) for the signed-Bernoulli distribution. Note that geometric weights require $W \geq 2$ and signed-Bernoulli weights require $0 < W \leq 2$, explaining the limited range of their corresponding curves. (b)  Detectability threshold $\Delta_{w}^*$ for the weighted planted partition model as a function of $W$, with the same parameters as in (a) but for the case of homogeneous topology, given by Eq.~(\ref{eq:poisson_w_det})
    for the Poisson distribution, Eq.~(\ref{eq:dirac_w_det}) for the Dirac distribution, Eq.~(\ref{eq:expon_w_det}) for the exponential distribution, Eq.~(\ref{eq:geometric_w_det}) for the geometric distribution, and Eq.~(\ref{eq:sign_w_det}) for the signed-Bernoulli distribution. The limited range of geometric and signed-Bernoulli weight curves reflects their respective parameter constraints.}
    \label{fig:1}
\end{figure}

In Figure~\ref{fig:1}(a), we plot the predictions of the detectability thresholds of Eqs.~(\ref{eq:dirac_det}), ~(\ref{eq:poisson_det}),~(\ref{eq:expon_det}), ~(\ref{eq:geometric_det}) and~(\ref{eq:sign_det})  as functions of $W$. Here $N=1024$ and $K=128$.
We note that limited ranges for some curves are shown due to parameter constraints: for Poisson weights require $W \geq 0$, exponential weights require $W > 0$, geometric weights require $W \geq 2$, and signed-Bernoulli weights require $0 < W \leq 2$; while Dirac weights have no restrictions.
Theory predicts that the easiest networks for the detection of communities  have weights obeying the Dirac distribution. Communities are easier to be detected in network with exponentially distributed weights than in networks with Poisson distributed weights for small $W$ values.
However, as $W$ grows, the detectability in networks for Poisson distributed weights decreases, eventually becoming smaller than in networks with exponentially distributed weights. In the limit of large $W$, the detectability threshold in networks with Poisson distributed weights approaches the one valid for networks with Dirac distributed weights.
The detectability threshold for exponential edge weights instead remains larger than in the other two cases even in the limit of infinitely large $W$ values.

\begin{figure*}[!htb]
    \centering
    \includegraphics[width=0.8\textwidth]{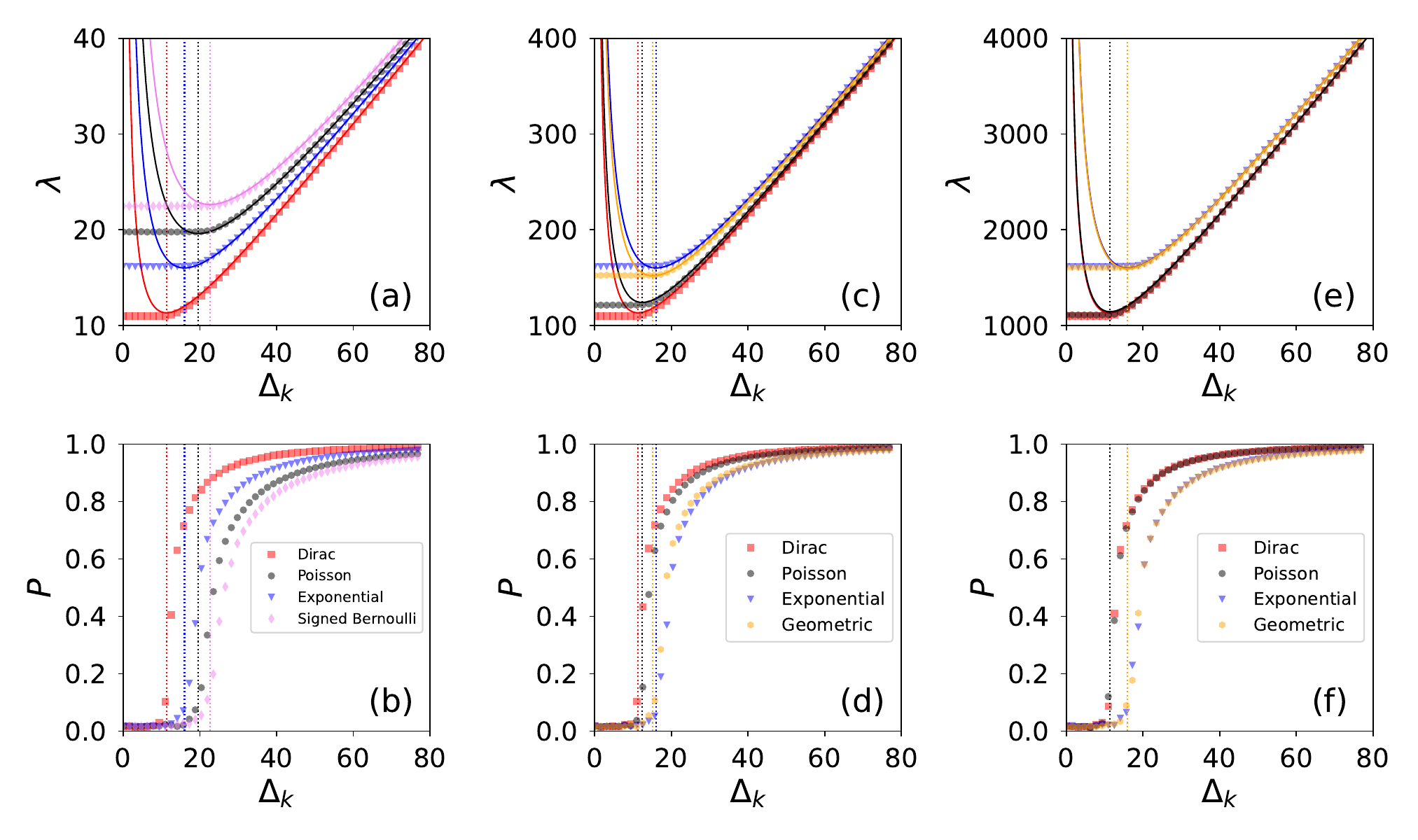}
    \caption{(a) Largest eigenvalue $\lambda$ of the modularity matrix as a function of the difference between the expected value of the within- and cross-community degrees $\Delta_k$  for the weighted planted-partition model. Here, $N=1024$, $K=128$, $W = 1$.
    Different symbols/colors correspond to different distributions of the edge weights.
    Numerical results are averaged over $100$ realizations of the model for each $\Delta_k$ value and displayed with symbols. The solid curves represent the theoretical predictions, whereas the vertical dotted lines denote the detectability thresholds. Note that some distributions are not shown for certain $W$ values due to parameter compatibility constraints (e.g., geometric weights require $W \geq 2$, signed-Bernoulli weights require $0 < W \leq 2$). (b) We plot the order parameter $P$ (i.e., Eq.~(\ref{eq:order})) as a function of $\Delta_k$ for the same networks as in (a). (c) Same as in (a), but for $W = 10$.  (d) Same as in (b), but for $W = 10$. (e) Same as in (a), but for $W = 100$.  (f) Same as in (b), but for $W = 100$.}
    \label{fig:2}
\end{figure*}

The above predictions are validated in numerical experiments conducted on instances of the weighted PPM. Results are reported in Figure~\ref{fig:2}. First of all, the behavior of the largest eigenvalue $\lambda$ of the modularity matrix as a function of $W$ is very well predicted by our Eqs.~(\ref{eq:lambda_dirac}), ~(\ref{eq:lambda_poisson}), and~(\ref{eq:lambda_expon}), see Figure~\ref{fig:2}(a), (c), and (e). Second the detection of the ground-truth community structure is possible only when the difference between within- and cross-cluster degrees exceeds the detectability thresholds of Eqs.~(\ref{eq:dirac_det}),~(\ref{eq:poisson_det}) and~(\ref{eq:expon_det}), see Figure~\ref{fig:2}(b), (d), and (f).

\begin{figure*}[!htb]
    \centering
    \includegraphics[width=0.8\textwidth]{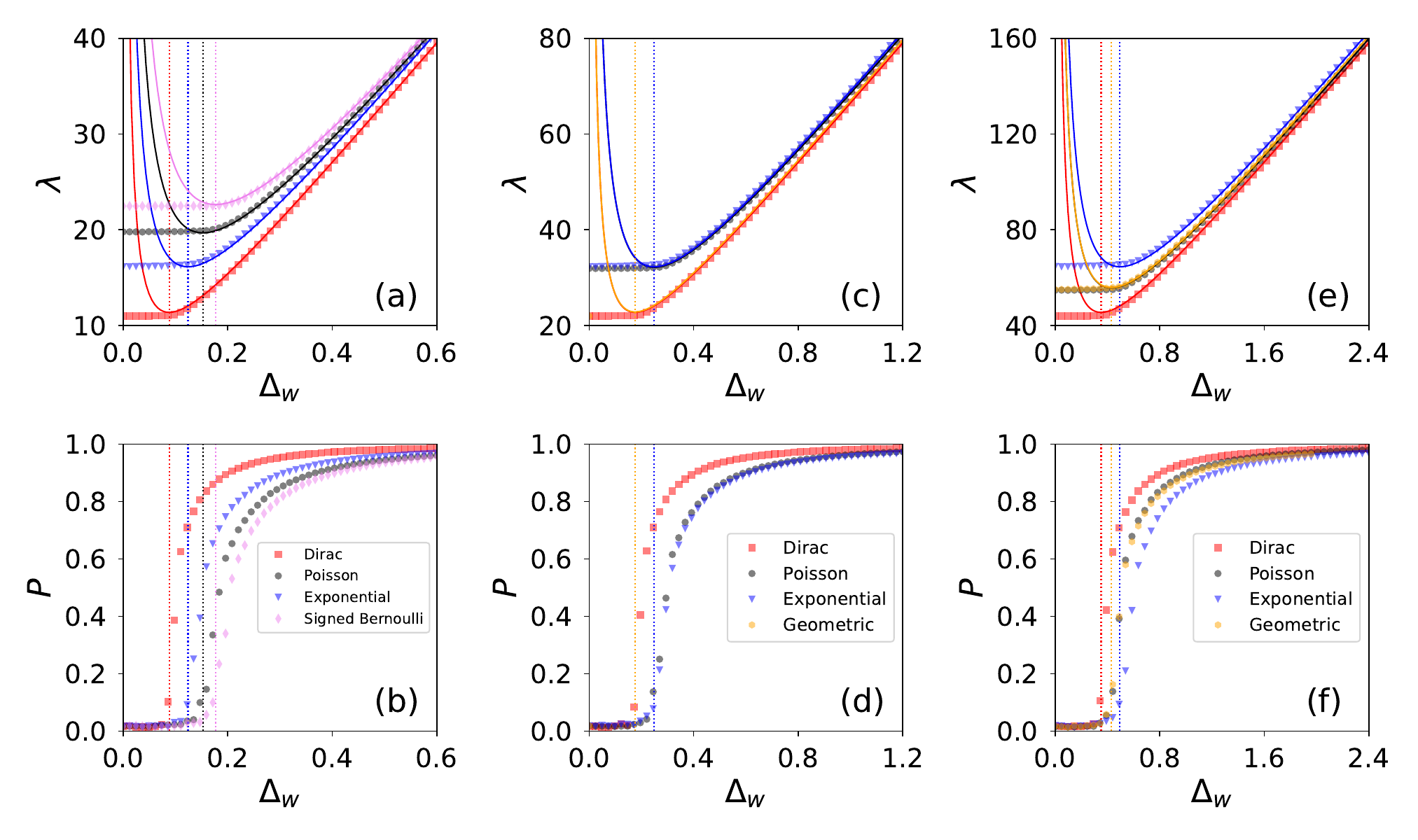}
    \caption{(a) Largest eigenvalue $\lambda$ of the modularity matrix as a function of the difference between the expected value of the within- and cross-community edge weights $\Delta_{w}$ for the weighted planted-partition model. Here, $N=1024$, $K=128$, $W = 1$. Different symbols/colors correspond to different distributions of the edge weights.
    Numerical results are averaged over $100$ realizations of the model for each $\Delta_{w}$ value and displayed with symbols. The solid curves represent the theoretical predictions, whereas the vertical dotted lines denote the detectability thresholds. Missing curves for certain $W$ values reflect parameter compatibility constraints (e.g., geometric weights require $W \geq 2$, while signed-Bernoulli weights are limited to $0 < W \leq 2$). (b) We plot the order parameter $P$ [i.e., Eq.~(\ref{eq:order})] as a function of $\Delta_w$ for the same networks as in (a). (c) Same as in (a), but for $W = 2$.  (d) Same as in (b), but for $W = 2$. (e) Same as in (a), but for $W = 4$.  (f) Same as in (b), but for $W = 4$.}
    \label{fig:3}
\end{figure*}

Similar considerations are valid when the ground-truth community structure is represented by edge weights only, whereas the underlying topology of the graph is completely independent of the ground-truth partition. Also in this case, we predict an inversion in the difficulty of retrieving the ground-truth partition between the cases of Poisson and exponential edge weights, see Figure~\ref{fig:1}(b), and such a prediction is validated in numerical experiments, see Figure~\ref{fig:3}.

\section*{Discussion}

This paper extends
the theory of community detectability to weighted networks by analyzing a weighted variant of the planted-partition model where edge weights follow different probability distributions~\cite{aicher2014learning,peixoto2018nonparametric}. Our key finding is that the detectability threshold depends non-trivially on both the topology and the edge weights, with the second moment of edge weight distribution playing a pivotal role in determining the difficulty of the community-detection task.

Our unified theoretical framework reveals a clear hierarchy in the difficulty of detecting communities across different weight distributions, with networks having Dirac-distributed weights (constant weights within and between communities) providing the easiest scenario for detection, achieving the smallest possible detectability threshold $\Delta_k^* = \sqrt{K}$. 
This result validates our theoretical framework by recovering the classical unweighted case.

In contrast, exponentially distributed weights consistently present a more challenging scenario, with a detectability threshold that is exactly $\sqrt{2}$ times larger than the Dirac case, regardless of the total average weight $W$. This $\sqrt{2}$ penalty factor reflects the high variance inherent in exponential distributions, which adds noise that obscures community structure. Poisson-distributed weights exhibit an intermediate and weight-dependent behavior, being more challenging than exponential distributions for small total weights but asymptotically approaching the  Dirac case as 
the average value of the weights increases.
This transition occurs because the relative variance of Poisson distributions decreases as their mean increases, effectively reducing the noise that hinders detection. The geometric and signed-Bernoulli weight distributions provide additional insights into the role of variance and support constraints, with geometric weights falling between the Poisson and the exponential cases and approaching exponential behavior as 
average value of the edges' weights increases,
while the signed-Bernoulli weights present a unique scenario where detectability depends inversely on the sign-imbalance parameter.

Our results establish that the detectability threshold in weighted networks is fundamentally determined by how variance scales with the mean across different weight distributions. Poisson weights exhibit linear scaling where variance increases proportionally with the mean, allowing detectability to approach the optimal Dirac case for large weights. Geometric weights show quadratic scaling where variance increases faster than the mean, making thresholds approach the challenging exponential case.
This result suggests that reducing the variance of edge weights relative to the mean edge weight can improve the community detectability for the PPM.
For instance, logarithmic transformations and rank-based methods may reduce the influence of excessively large weights, reducing the variance.
We leave for future work the systematic evaluation of preprocessing techniques that affect detectability thresholds across various weight distributions and network topologies.

Despite these insights, our analysis relies on several important assumptions that limit its generalizability. We consider only two equally sized communities, while real networks often contain multiple communities of varying sizes. To address this limitation, we extended our analysis to multi-community networks using the Leiden algorithm, which confirmed numerically that the same hierarchy of weight distributions holds across different numbers of communities (Appendix~\ref{sec:more_communities}). However, the extension to more complex community structures with varying sizes remains an open challenge, though our unified framework provides a foundation for such investigations. 
Also, our model assumes 
conditional independence between the generation of the edges and their weights given the community partition,
but in practice there may be correlations between node degrees and the weights of their connections, which could alter the detectability thresholds. This represents an important direction for future work. Last but not least, we focus exclusively on spectral modularity optimization for community detection, and while our theoretical predictions align with information-theoretic limits, validating these results across different detection algorithms would strengthen the generalizability of our findings.
\change{We expect that the same detectability limits apply to other modularity-maximization methods, as evidenced by our numerical validation using the Leiden algorithm (Appendix D). More broadly, while exact numerical thresholds may vary across different community detection methods, we expect the hierarchy among weight distributions to remain consistent. This hierarchy is fundamental because it reflects the signal-to-noise ratio inherent in weighted networks: signal arises from differences in average edge weights between and within communities, while noise stems from the random fluctuations of edge weights characterized by the variance. Thus, we believe that Dirac weights are consistently the easiest case for community detection, exponential weights are more challenging, and Poisson weights exhibit intermediate behavior for other community detection methods.} 



\subsection*{Acknowledgement} \label{sec:ack}
This project was partially supported by the Air Force Office of Scientific Research under award numbers FA9550-19-1-0391, FA9550-21-1-0446 and FA9550-24-1-0039, and by the National Science Foundation under award numbers 1927418, and by the National Institutes of Health under awards U01 AG072177 and U19 AG074879.


%

\appendix

\section{Detailed calculation of MGF derivatives}
\label{sec:moment_calculations}

We provide the calculations for obtaining the first and second moments of the distribution ${\cal D}$ using moment generating functions.
The first and second moments of the distribution ${\cal D}$ are obtained as $\langle \Delta_{s} \rangle = M'_{\cal D}(0)$ and $\langle \Delta_{s}^2 \rangle = M''_{\cal D}(0)$, respectively, where for compactness we denote $M'(x) = \frac{\partial}{\partial x} M(x)$ and $M''(x) = \frac{\partial^2}{\partial x^2} M(x)$.

The first derivative of $M_{\cal D}(x)$ is:
\[
M'_{\cal D}(x) = M'_{{\cal S}_\text{in}}(x) M_{{\cal S}_\text{out}}(-x) - M'_{{\cal S}_\text{out}}(-x) M_{{\cal S}_\text{in}}(x),
\]
and the second derivative is:
\[
\begin{array}{ll}
M''_{\cal D}(x) = & M''_{{\cal S}_\text{in}}(x) M_{{\cal S}_\text{out}}(-x) - 2 M'_{{\cal S}_\text{in}}(x) M'_{{\cal S}_\text{out}}(-x) \\
& + M''_{{\cal S}_\text{out}}(-x) M_{{\cal S}_\text{in}}(x).
\end{array}
\]

The first derivative of $M_{{\cal S}_\text{in}}(x)$ is:
\[
M'_{{\cal S}_\text{in}}(x) = M'_{{\cal W}_\text{in}}(x) \sum_{k} {\cal K}_\text{in}(k) k \left[ M_{{\cal W}_\text{in}}(x) \right]^{k-1},
\]
and the second derivative is:
\[
\begin{array}{ll}
M''_{{\cal S}_\text{in}}(x) = & M''_{{\cal W}_\text{in}}(x) \sum_{k} {\cal K}_\text{in}(k) k \left[ M_{{\cal W}_\text{in}}(x) \right]^{k-1} \\
& + [M'_{{\cal W}_\text{in}}(x)]^2 \sum_{k} {\cal K}_\text{in}(k) k(k-1) \left[ M_{{\cal W}_\text{in}}(x) \right]^{k-2}.
\end{array}
\]
Similar expressions hold for the derivatives of $M_{{\cal S}_\text{out}}(x)$.

Now we provide the detailed step-by-step derivations for equations~\eqref{eq:mS1}, \eqref{eq:mS2}, and \eqref{eq:mD1}.

\subsection*{Step-by-step derivation of Eq.~\eqref{eq:mS1}}

We start with the MGF of the within-cluster strength distribution:
\[
M_{{\cal S}_\text{in}}(x) = \sum_{k} {\cal K}_\text{in}(k) \left[ M_{{\cal W}_\text{in}}(x) \right]^k
\]

Taking the first derivative with respect to $x$:
\[
M'_{{\cal S}_\text{in}}(x) = \sum_{k} {\cal K}_\text{in}(k) \frac{d}{dx}\left[ M_{{\cal W}_\text{in}}(x) \right]^k
\]


Therefore:
\[
M'_{{\cal S}_\text{in}}(x) = M'_{{\cal W}_\text{in}}(x) \sum_{k} {\cal K}_\text{in}(k) k \left[ M_{{\cal W}_\text{in}}(x) \right]^{k-1}
\]

Evaluating at $x = 0$ and using $M_{{\cal W}_\text{in}}(0) = 1$ and $M'_{{\cal W}_\text{in}}(0) = \langle w_\text{in} \rangle$:
\[
M'_{{\cal S}_\text{in}}(0) = \langle w_\text{in} \rangle \sum_{k} {\cal K}_\text{in}(k) k \cdot 1^{k-1} = \langle w_\text{in} \rangle \sum_{k} {\cal K}_\text{in}(k) k
\]

Since $\sum_{k} {\cal K}_\text{in}(k) k = \langle k_{\text{in}} \rangle$ by definition of expectation:
\[
\langle s_\text{in} \rangle = M'_{{\cal S}_\text{in}}(0) = \langle w_\text{in} \rangle \langle k_{\text{in}} \rangle
\]

This gives us Eq.~\eqref{eq:mS1}.

\subsection*{Step-by-step derivation of Eq.~\eqref{eq:mS2}}

For the second moment, we need the second derivative of $M_{{\cal S}_\text{in}}(x)$. Starting from:
\[
M'_{{\cal S}_\text{in}}(x) = M'_{{\cal W}_\text{in}}(x) \sum_{k} {\cal K}_\text{in}(k) k \left[ M_{{\cal W}_\text{in}}(x) \right]^{k-1}
\]

Taking the derivative again using the product rule:
\[
\begin{array}{ll}
M''_{{\cal S}_\text{in}}(x) = & M''_{{\cal W}_\text{in}}(x) \sum_{k} {\cal K}_\text{in}(k) k \left[ M_{{\cal W}_\text{in}}(x) \right]^{k-1} \\
& + \left[M'_{{\cal W}_\text{in}}(x)\right]^2 \sum_{k} {\cal K}_\text{in}(k) k(k-1) \left[ M_{{\cal W}_\text{in}}(x) \right]^{k-2}
\end{array}
\]


Evaluating at $x = 0$ using $M_{{\cal W}_\text{in}}(0) = 1$, $M'_{{\cal W}_\text{in}}(0) = \langle w_\text{in} \rangle$, and $M''_{{\cal W}_\text{in}}(0) = \langle w^2_\text{in} \rangle$:
\[
\begin{array}{ll}
M''_{{\cal S}_\text{in}}(0) = & \langle w^2_\text{in} \rangle \sum_{k} {\cal K}_\text{in}(k) k \\
& + \langle w_\text{in} \rangle^2 \sum_{k} {\cal K}_\text{in}(k) k(k-1)
\end{array}
\]

Since $\sum_{k} {\cal K}_\text{in}(k) k = \langle k_{\text{in}} \rangle$ and $\sum_{k} {\cal K}_\text{in}(k) k(k-1) = \langle k^2_{\text{in}} \rangle - \langle k_{\text{in}} \rangle$:
\[
\langle s^2_\text{in} \rangle = M''_{{\cal S}_\text{in}}(0) = \langle w^2_\text{in} \rangle \langle k_{\text{in}} \rangle + \langle w_\text{in} \rangle^2 \left( \langle k^2_{\text{in}} \rangle - \langle k_{\text{in}} \rangle \right)
\]

This gives us Eq.~\eqref{eq:mS2}.

\subsection*{Step-by-step derivation of Eq.~\eqref{eq:mD1}}

The MGF of the strength difference distribution is:
\[
M_{\cal D}(x) = \langle e^{x (s_\text{in} - s_\text{out})} \rangle = M_{{\cal S}_\text{in}}(x) M_{{\cal S}_\text{out}}(-x)
\]

Taking the first derivative:
\[
M'_{\cal D}(x) = M'_{{\cal S}_\text{in}}(x) M_{{\cal S}_\text{out}}(-x) - M'_{{\cal S}_\text{out}}(-x) M_{{\cal S}_\text{in}}(x)
\]


Evaluating at $x = 0$ and using $M_{{\cal S}_\text{in}}(0) = M_{{\cal S}_\text{out}}(0) = 1$:
\[
M'_{\cal D}(0) = M'_{{\cal S}_\text{in}}(0) - M'_{{\cal S}_\text{out}}(0)
\]

Since $M'_{{\cal S}_\text{in}}(0) = \langle s_\text{in} \rangle$ and $M'_{{\cal S}_\text{out}}(0) = \langle s_\text{out} \rangle$:
\[
\langle \Delta_s \rangle = M'_{\cal D}(0) = \langle s_\text{in} \rangle - \langle s_\text{out} \rangle
\]

This gives us Eq.~\eqref{eq:mD1}.

\subsection*{Step-by-step derivation of Eq.~\eqref{eq:mD2}}

For the second moment of the strength difference, we need the second derivative of $M_{\cal D}(x)$. Starting from:
\[
M'_{\cal D}(x) = M'_{{\cal S}_\text{in}}(x) M_{{\cal S}_\text{out}}(-x) - M'_{{\cal S}_\text{out}}(-x) M_{{\cal S}_\text{in}}(x)
\]

Taking the derivative again using the product rule:
\[
\begin{array}{ll}
M''_{\cal D}(x) = & M''_{{\cal S}_\text{in}}(x) M_{{\cal S}_\text{out}}(-x) - 2 M'_{{\cal S}_\text{in}}(x) M'_{{\cal S}_\text{out}}(-x) \\
& + M''_{{\cal S}_\text{out}}(-x) M_{{\cal S}_\text{in}}(x)
\end{array}
\]


Evaluating at $x = 0$ and using $M_{{\cal S}_\text{in}}(0) = M_{{\cal S}_\text{out}}(0) = 1$:
\[
M''_{\cal D}(0) = M''_{{\cal S}_\text{in}}(0) - 2 M'_{{\cal S}_\text{in}}(0) M'_{{\cal S}_\text{out}}(0) + M''_{{\cal S}_\text{out}}(0)
\]

Since $M'_{{\cal S}_\text{in}}(0) = \langle s_\text{in} \rangle$, $M'_{{\cal S}_\text{out}}(0) = \langle s_\text{out} \rangle$, $M''_{{\cal S}_\text{in}}(0) = \langle s^2_\text{in} \rangle$, and $M''_{{\cal S}_\text{out}}(0) = \langle s^2_\text{out} \rangle$:
\[
\langle \Delta_s^2 \rangle = M''_{\cal D}(0) = \langle s^2_\text{in} \rangle - 2 \langle s_\text{in} \rangle \langle s_\text{out} \rangle + \langle s^2_\text{out} \rangle
\]

This gives us Eq.~\eqref{eq:mD2}.


\section{Detailed calculation of largest eigenvalues and detectability thresholds}
\label{sec:threshold_calculations}

\subsection*{Step-by-step derivation of Eq.~\eqref{eq:main_equation}}
\label{sec:derivation_of_main_equation}

Let us derive $\lambda$ as a function of the moments starting from Eq.~\eqref{eq:main}.
In particular, we have:
\begin{align*}
\langle \Delta_{s} \rangle &= \langle s_\text{in} \rangle - \langle s_\text{out} \rangle, \\
\langle s_\text{in} \rangle &= \langle w_\text{in} \rangle \langle k_{\text{in}} \rangle, \\
\langle s_\text{out} \rangle &= \langle w_\text{out} \rangle \langle k_{\text{out}} \rangle.
\end{align*}

The second moment (Eq.~\eqref{eq:mD2}) is given by:
\begin{equation*}
\langle \Delta_{s}^2 \rangle = \langle s^2_\text{in} \rangle - 2 \langle s_\text{in} \rangle \langle s_\text{out} \rangle + \langle s^2_\text{out} \rangle,
\end{equation*}
where
\begin{align*}
\langle s^2_\text{in} \rangle &= \langle w^2_\text{in} \rangle \langle k_{\text{in}} \rangle + \langle w_\text{in} \rangle^2 (\langle k_{\text{in}}^2 \rangle - \langle k_{\text{in}} \rangle), \\
\langle s^2_\text{out} \rangle &= \langle w^2_\text{out} \rangle \langle k_{\text{out}} \rangle + \langle w_\text{out} \rangle^2 (\langle k_{\text{out}}^2 \rangle - \langle k_{\text{out}} \rangle).
\end{align*}


Finally, substituting into the eigenvalue formula and simplifying:
\begin{align*}
\lambda &= \frac{\langle k_{\text{in}} \rangle \langle w^2_\text{in} \rangle + \langle k_{\text{out}} \rangle \langle w^2_\text{out} \rangle}{\langle s_\text{in} \rangle - \langle s_\text{out} \rangle} + (\langle s_\text{in} \rangle - \langle s_\text{out} \rangle)
\end{align*}

Substituting $\langle s_\text{in} \rangle = \langle k_{\text{in}} \rangle \langle w_\text{in} \rangle$ and $\langle s_\text{out} \rangle = \langle k_{\text{out}} \rangle \langle w_\text{out} \rangle$, we get Eq.~\eqref{eq:main_equation}.

\subsection*{Step-by-step derivation of Eq.~\eqref{eq:lambda_moment_homogeneous_weights}}
\label{sec:derivation_of_lambda}
\label{sec:detailed_derivation_homogeneous}

For homogeneous weights across communities, we have:
\begin{align*}
\langle w_\text{in} \rangle &= \langle w_\text{out} \rangle = \frac{W}{2},\\
\Delta_{w} &= \langle w_\text{in} \rangle - \langle w_\text{out} \rangle = 0,
\end{align*}
where $W$ is defined in Eq.~(\ref{eq:omega}). This condition ensures that community structure manifests only through network topology, not through weight heterogeneity.

We express the intra-community degree and inter-community degree in terms of the total degree $K$ and degree difference $\Delta_k$:
\begin{align*}
\langle k_{\text{in}} \rangle &= \frac{K + \Delta_k}{2},\\
\langle k_{\text{out}} \rangle &= \frac{K - \Delta_k}{2},
\end{align*}
where $K = \langle k_{\text{in}} \rangle + \langle k_{\text{out}} \rangle$ and $\Delta_k = \langle k_{\text{in}} \rangle - \langle k_{\text{out}} \rangle$.

For the second-order polynomial family of weight distributions (Eq.~\eqref{eq:second_order_polynomial}), the second moments are:
\begin{align*}
\langle w^2_\text{in} \rangle = \langle w^2_\text{out} \rangle = \alpha_0 + \frac{\alpha_1 W}{2} + \frac{\alpha_2 W^2}{4}
\end{align*}

Looking now at Eq.~\eqref{eq:main_equation}, the denominator becomes:
\begin{align*}
\text{Denominator} &= \langle k_{\text{in}} \rangle \langle w_\text{in} \rangle - \langle k_{\text{out}} \rangle \langle w_\text{out} \rangle = \frac{\Delta_k W}{2}
\end{align*}
Since $\langle w^2_\text{in} \rangle = \langle w^2_\text{out} \rangle$ under homogeneous conditions, the numerator becomes:
\begin{align*}
\text{Numerator} &= (\langle k_{\text{in}} \rangle + \langle k_{\text{out}} \rangle) \langle w^2_\text{in} \rangle = K \left(\alpha_0 + \frac{\alpha_1 W}{2} + \frac{\alpha_2 W^2}{4}\right)
\end{align*}


To simplify the expression, we define $C = 4\alpha_0 + 2W\alpha_1 + W^2\alpha_2$, so that:
\begin{align*}
\lambda &= \frac{\Delta_k}{2}W + \frac{1}{\Delta_k} \cdot \frac{KC}{2W}
\end{align*}

We have thus derived the eigenvalue expression for homogeneous weights as Eq.~\eqref{eq:lambda_moment_homogeneous_weights}.

\subsection*{Step-by-step derivation of Eq.~\eqref{eq:lambda_moment_homogeneous_weights_general_omega}}
\label{sec:detailed_derivation_homogeneous_topology}

We provide a derivation of the eigenvalue expression and detectability threshold for the complementary case where network topology is homogeneous (i.e., $\langle k_{\text{in}} \rangle = \langle k_{\text{out}} \rangle = K/2$), but communities are encoded solely through edge weight differences. This corresponds to an Erd\H{o}s--R\'enyi random graph where community structure manifests only through weight heterogeneity.

For homogeneous topology, we have:
\begin{align*}
\langle k_{\text{in}} \rangle &= \langle k_{\text{out}} \rangle = \frac{K}{2},\\
\Delta_k &= \langle k_{\text{in}} \rangle - \langle k_{\text{out}} \rangle = 0,
\end{align*}
where community structure is encoded through weight differences:
\begin{align*}
\langle w_\text{in} \rangle &= \frac{W}{2} + \frac{\Delta_{w}}{2},\\
\langle w_\text{out} \rangle &= \frac{W}{2} - \frac{\Delta_{w}}{2},
\end{align*}
where $\Delta_{w} = \langle w_\text{in} \rangle - \langle w_\text{out} \rangle$ is the weight difference between communities and $W = \langle w_\text{in} \rangle + \langle w_\text{out} \rangle$ is the total average weight.

For the second-order polynomial family of weight distributions, the second moments are:
\begin{align*}
\langle w^2_\text{in} \rangle &= \alpha_0 + \alpha_1 \frac{W + \Delta_{w}}{2} + \alpha_2 \frac{(W + \Delta_{w})^2}{4},\\
\langle w^2_\text{out} \rangle &= \alpha_0 + \alpha_1 \frac{W - \Delta_{w}}{2} + \alpha_2 \frac{(W - \Delta_{w})^2}{4}.
\end{align*}

Starting from the general eigenvalue formula (Eq.~\eqref{eq:main_equation}), the denominator becomes:
\begin{align*}
\text{Denominator} &= \langle k_{\text{in}} \rangle \langle w_\text{in} \rangle - \langle k_{\text{out}} \rangle \langle w_\text{out} \rangle = \frac{K\Delta_{w}}{2}
\end{align*}

For the numerator, we compute:
\begin{align*}
\text{Numerator} &= \frac{K}{2} \left(\langle w^2_\text{in} \rangle + \langle w^2_\text{out} \rangle\right)
\end{align*}


We can rewrite this in terms of the parameter $C = 4\alpha_0 + 2W\alpha_1 + W^2\alpha_2$. Therefore, the eigenvalue expression becomes:
\begin{align*}
\lambda &= \frac{\Delta_{w}}{2}(K + \alpha_2) + \frac{C}{2\Delta_{w}}
\end{align*}

To find the detectability threshold, we take the derivative with respect to $\Delta_{w}$ and set it equal to zero:
\begin{align*}
\frac{\partial \lambda}{\partial \Delta_{w}} &= \frac{K + \alpha_2}{2} - \frac{C}{2\Delta_{w}^2} = 0,
\end{align*}
which gives:
\begin{align*}
\Delta_{w}^* &= \sqrt{\frac{C}{K + \alpha_2}} = \sqrt{\frac{4\alpha_0 + 2W\alpha_1 + W^2\alpha_2}{K + \alpha_2}}
\end{align*}

We have thus derived the detectability threshold for homogeneous topology given by Eq.~\eqref{eq:delta_omega_threshold_homogeneous_topology}.

The eigenvalue expression for homogeneous topology reveals that detectability depends on the competition between a linear term in $\Delta_{w}$ (scaled by $K + \alpha_2$) and an inverse term containing the weight distribution variance parameter $C$. Unlike the homogeneous weight case, the coefficient of the linear term includes $\alpha_2$, reflecting the additional contribution from weight variance when weights are heterogeneous across communities.

\section{Correlation matrix}
\label{sec:correlation_matrix}

Many weighted networks are constructed based on correlation matrices. To align this case with the unified framework established in the main text, we show how correlation matrices fit within the second-order polynomial parameterization of Eq.~\eqref{eq:second_order_polynomial}.

It is important to note that correlation matrices fundamentally violate the conditional independence assumption underlying our theoretical framework. While our general approach assumes that edge weights are generated independently given the community assignments, correlation coefficients between different pairs of variables are inherently dependent on each other due to the shared underlying data matrix. Despite this violation, our theoretical results closely aligned with the numerical results, as demonstrated below.

We consider $N$-dimensional gaussian random vectors with zero mean, unit variance, and $L$ observations:
\begin{equation}
\mathbf{X} = \left[ \mathbf{x}_1, \mathbf{x}_2, \ldots, \mathbf{x}_L \right] \; ,
\end{equation}
where $\mathbf{x}_i$ is the $i$-th observation of the $N$-dimensional gaussian random vector. The correlation matrix is:
\begin{equation}
    \mathbf{R} = \frac{1}{L} \mathbf{X} \mathbf{X}^T \; .
\end{equation}

Before considering the detectability limit, let us prepare ourselves by identifying the distribution of the sample correlation coefficient $r$.
The exact probability density function of the sample correlation coefficient $r$ is rather complex. Thus, we use an approximation based on the Fisher-$z$ transform, which is accurate when the sample size $L$ is large. Specifically, for a sample correlation coefficient $r$, we apply the Fisher-$z$ transform:
\begin{equation}
    z(r) := \text{tanh}^{-1}(r) = \frac{1}{2} \log \left( \frac{1+r}{1-r} \right) \; ,
\end{equation}
where $r$ is the sample correlation coefficient.
For a large sample size $L$, $z(r)$ follows a gaussian distribution with mean being the true correlation $\rho$ and variance $\frac{1}{L-3}$~\cite{hotelling1953new}.
Taking the Taylor expansion of $r=\text{tanh}(z(r))$ around $\rho$ leads to
\begin{align}
    \text{tanh}(z(\rho) + \epsilon) &\approx \text{tanh}(z(\rho)) + (1-\text{tanh}^2(z(\rho))) \epsilon + \mathcal{O}(\epsilon^2) \nonumber \\
    &=\rho + (1-\rho^2) \epsilon + \mathcal{O}(\epsilon^2),
\end{align}
Thus the sample correlation follows approximately a gaussian distribution~\cite{hotelling1953new}, i.e.,
\begin{equation}
    R_{ij} \sim {\cal N}\left(\rho, \frac{(1-\rho^2_{ij})^2}{L-3}\right). \label{eq:hotelling_dist}
\end{equation}

Using this distribution, the
second moment is
\begin{align}
    \langle w_\text{in}^2 \rangle &= \langle w_\text{in} \rangle^2 +  \frac{1}{L-3} \left(1 - \langle w_\text{in} \rangle^2\right)^2
\end{align}
and similar for the cross-cluster weights.
Plugging these into Eq.~(\ref{eq:main_equation}), we have:
\begin{align}
    \lambda &=\langle k_{\text{in}} \rangle \langle w_\text{in} \rangle - \langle k_{\text{out}} \rangle \langle w_\text{out} \rangle \nonumber\\
    &\quad +  \frac{\langle k_{\text{in}} \rangle \left(\langle w_\text{in} \rangle^{2} + \frac{\left(1 - \langle w_\text{in} \rangle^{2}\right)^{2}}{L - 3}\right) + \langle k_{\text{out}} \rangle \left(\langle w_\text{out} \rangle^{2} + \frac{\left(1 - \langle w_\text{out} \rangle^{2}\right)^{2}}{L - 3}\right)}{\langle k_{\text{in}} \rangle \langle w_\text{in} \rangle - \langle k_{\text{out}} \rangle \langle w_\text{out} \rangle}
\end{align}
Assuming $L \gg 1$, we have
\begin{align}
    \lambda =\langle k_{\text{in}} \rangle \langle w_\text{in} \rangle - \langle k_{\text{out}} \rangle \langle w_\text{out} \rangle +  \frac{\langle k_{\text{in}} \rangle \langle w_\text{in} \rangle^{2} + \langle k_{\text{out}} \rangle \langle w_\text{out} \rangle^{2}}{\langle k_{\text{in}} \rangle \langle w_\text{in} \rangle - \langle k_{\text{out}} \rangle \langle w_\text{out} \rangle}
\end{align}
which reduces to the expression for Dirac-distributed weights [Eq.~\eqref{eq:lambda_dirac}]. This demonstrates that for large sample sizes, correlation-based networks behave similarly to networks with Dirac-distributed weights, as the finite sample size effects vanish and the sample correlation coefficients converge to their true values.
Expanding the second moment:
\begin{align}
    \langle w_\text{in}^2 \rangle &= \langle w_\text{in} \rangle^2 +  \frac{1}{L-3} \left(1 - 2\langle w_\text{in} \rangle^2 + \langle w_\text{in} \rangle^4\right) \\
    &= \frac{1}{L-3} + \langle w_\text{in} \rangle^2 \left(1 - \frac{2}{L-3}\right) + \frac{\langle w_\text{in} \rangle^4}{L-3}
\end{align}
For finite sample sizes, this corresponds to the polynomial parameters:
\begin{equation}
    (\alpha_0, \alpha_1, \alpha_2) = \left(\frac{1}{L-3}, -\frac{2}{L-3}, 1 + \frac{1}{L-3}\right)
    \label{eq:correlation_coefficients}
\end{equation}
Assuming a large sample size $L$ to ignore the higher-order terms, we have:
\begin{equation}
    (\alpha_0, \alpha_1, \alpha_2) \rightarrow (0, 0, 1)
\end{equation}
which corresponds exactly to the Dirac distribution case.

Using the unified framework equations from the main text, we can now apply Eqs.~\eqref{eq:lambda_moment_homogeneous_weights_general} and \eqref{eq:lambda_moment_homogeneous_weights_general_omega} directly. For the large sample limit, the detectability thresholds reduce to the Dirac case:
\begin{align}
    \Delta_k^* &= \sqrt{K} \quad \text{(homogeneous weights)}\\
    \Delta_{w}^* &= \frac{W}{\sqrt{K+1}} \quad \text{(homogeneous topology)}
\end{align}
This demonstrates that correlation-based networks behave as networks with Dirac-distributed weights in the large sample limit, providing a unified treatment within our general framework.

For finite sample sizes, we can substitute the correlation coefficients from Eq.~\eqref{eq:correlation_coefficients} into the general detectability expressions Eqs.~\eqref{eq:lambda_moment_homogeneous_weights_general} and \eqref{eq:lambda_moment_homogeneous_weights_general_omega} to obtain sample-size-dependent thresholds that smoothly transition to the Dirac case as $L$ increases.

\section{Networks with more than two communities}
\label{sec:more_communities}

We extend our analysis to networks with more than two communities. For each configuration of the PPM, we generate 100 network realizations with $N = 1024$ nodes and average degree $K = 128$.

Following previous work~\cite{kojaku2024network}, we specify $\langle k_{\text{in}} \rangle$ and $\langle k_{\text{out}} \rangle$ for $q$ communities such that the average degree is $K$:
\begin{align}
    K = \langle k_{\text{in}} \rangle + (q-1) \langle k_{\text{out}} \rangle
\end{align}
By substituting $\Delta_k = \langle k_{\text{in}} \rangle - \langle k_{\text{out}} \rangle$, we obtain:
\begin{align}
    \langle k_{\text{in}} \rangle &= \frac{K}{q} + \frac{q-1}{q} \Delta_k, \nonumber \\
    \langle k_{\text{out}} \rangle &= \frac{K}{q} - \frac{1}{q} \Delta_k. \nonumber
\end{align}
Similarly, we specify $\langle w_{\text{in}} \rangle$ and $\langle w_{\text{out}} \rangle$ for $q$ communities as:
\begin{align}
    \langle w_{\text{in}} \rangle &= \frac{W}{q} + \frac{q-1}{q} \Delta_w, \nonumber \\
    \langle w_{\text{out}} \rangle &= \frac{W}{q} - \frac{1}{q} \Delta_w. \nonumber
\end{align}

Edge weights $\langle w_{\text{in}} \rangle$ and $\langle w_{\text{out}} \rangle$ are functions of $(W, \Delta_w, q)$ and must fall within valid ranges for each distribution. Since the geometric and the signed-Bernoulli distributions have strictly bounded ranges that become incompatible for larger $q$, we focus on Dirac, exponential, and Poisson distributions to ensure fair comparisons with consistent parameter configurations across all cases.

While the spectral clustering approach provides theoretical insights for the two-community case, it lacks a principled extension to detect more than two communities. This limitation led us to use the Leiden algorithm~\cite{traag2019louvain}, which maximizes the same objective as the spectral clustering method but can detect an arbitrary number of communities. We measure similarity between detected and planted partitions using normalized mutual information (NMI).

\begin{figure*}[!htb]
    \centering
    \includegraphics[width=0.8\textwidth]{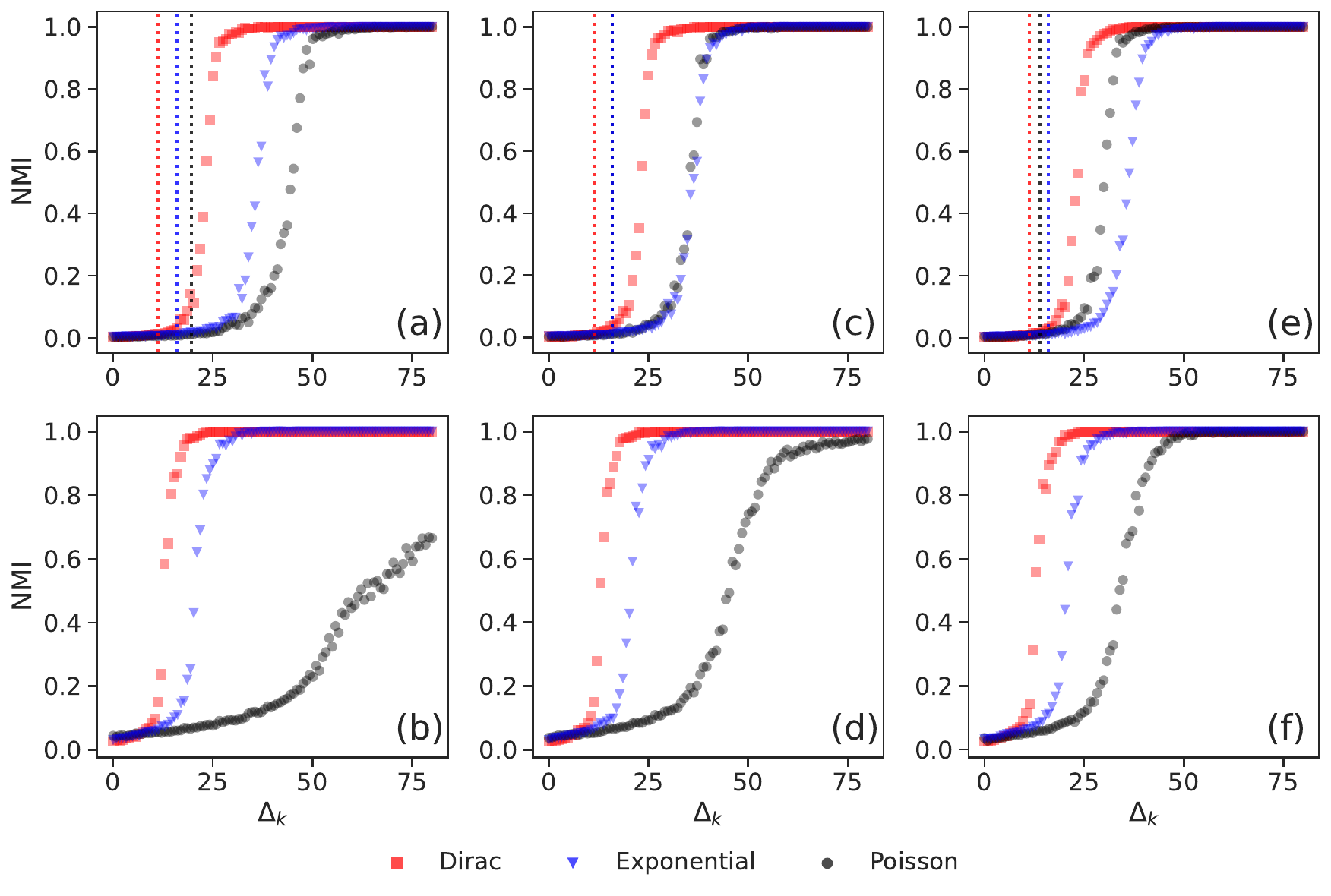}
    \caption{Community detection performance using the Leiden algorithm as a function of $\Delta_k$ for $\langle w_{\text{in}} \rangle = \langle w_{\text{out}} \rangle$. Top row: $q=2$ communities; bottom row: $q=16$ communities. Each line represents a different weight distribution. Networks have $N=1024$ nodes and average degree $K=128$. Panels show different total weights: (a,b) $W=1$; (c,d) $W=2$; (e,f) $W=4$.}
    \label{fig:leiden_delta_k}
\end{figure*}

\begin{figure*}
    \centering
    \includegraphics[width=0.8\textwidth]{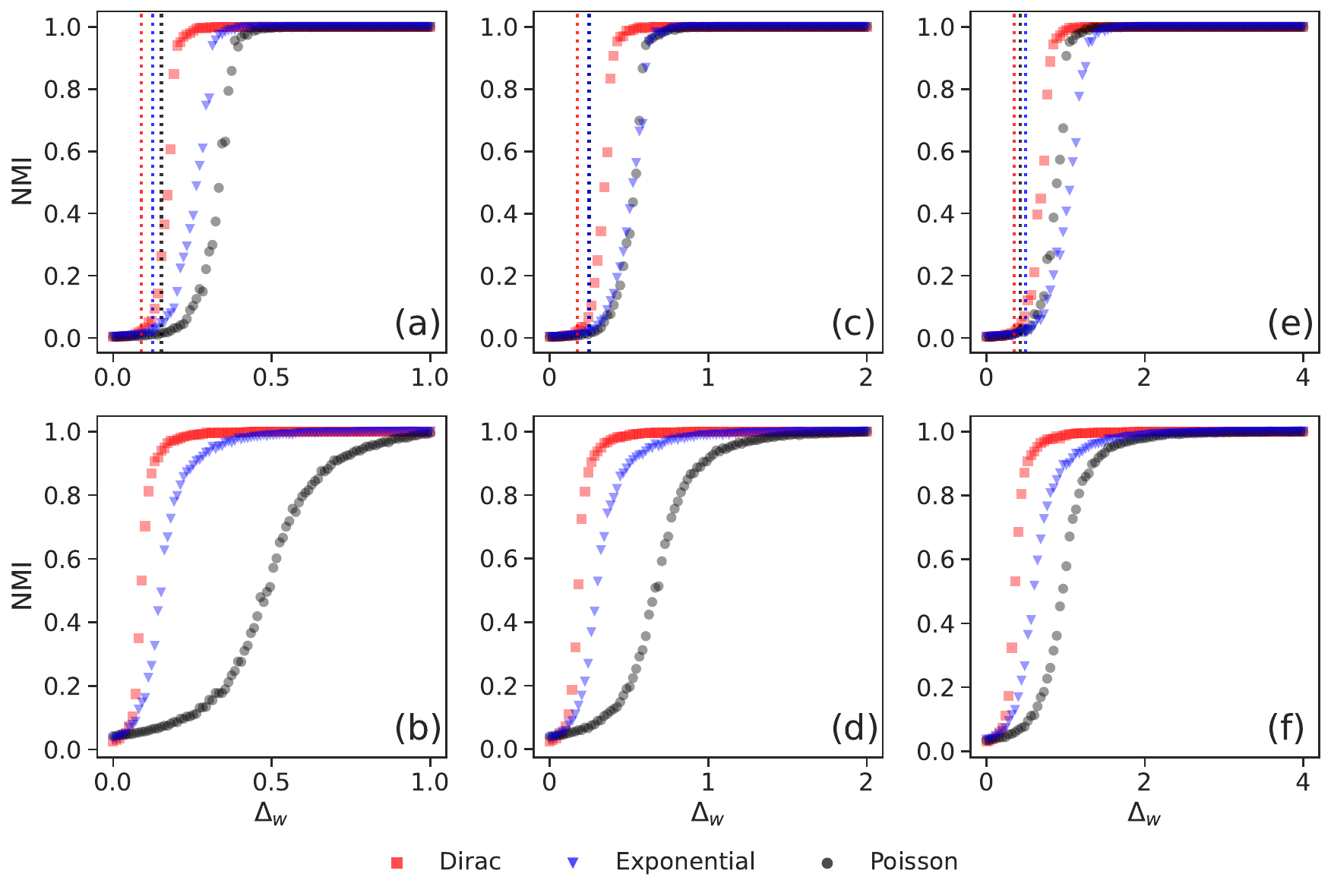}
    \caption{Community detection performance using the Leiden algorithm as a function of $\Delta_w$ for $\langle k_{\text{in}} \rangle = \langle k_{\text{out}} \rangle$. Top row: $q=2$ communities; bottom row: $q=16$ communities. Each line represents a different weight distribution. Networks have $N=1024$ nodes and average degree $K=128$. Panels show different total weights: (a,b) $W=1$; (c,d) $W=2$; (e,f) $W=4$.}
    \label{fig:leiden_delta_w}
\end{figure*}

Figures~\ref{fig:leiden_delta_k} and \ref{fig:leiden_delta_w} show NMI for the Leiden as functions of $\Delta_k$ and $\Delta_w$ for different weight distributions. The Leiden algorithm does not surpass the spectral clustering method nor bypass the theoretical detectability limit for the case of $q=2$ communities.
This performance gap may stem in part from the lack of knowledge of the true number of communities.
The Leiden algorithm freely selects any number of communities, while the spectral clustering method is constrained to the true number of communities. This constraint provides an advantage for spectral clustering. Importantly, even with knowledge of the true community number, spectral clustering does not bypass the detectability limit, consistent with our theoretical analysis.

The Leiden algorithm reproduces the same hierarchy of weight distributions observed in our spectral clustering analysis. Across all parameter configurations, Dirac weights exhibit the lowest detectability (easiest to detect communities), followed by Poisson weights, and exponential weights. Consistent with our theoretical predictions, increasing the total weight $W$ improves detectability for Poisson distributions, while the performance gap between Dirac and exponential distributions remains largely unchanged with varying $W$.

\end{document}